\documentclass[twocolumn,letterpaper,aps,prl,longbibliography,superscriptaddress,showpacs,floatfix]{revtex4-1}

\usepackage{graphicx}	% Include figure filed
\usepackage{xspace}	% Include xspace

\newcommand{\RdAu}{\mbox{$R_{d{\rm Au}}$}\xspace}
\newcommand{\RCP}{\mbox{$R_{\rm CP}$}\xspace}
\newcommand{\pT}{\mbox{$p_T$}\xspace}
\newcommand{\pTrec}{\mbox{$p_T^{\rm rec}$}\xspace}
\newcommand{\dAu}{\mbox{$d$$+$Au}\xspace}

\newcommand{\pp}{\mbox{$p$$+$$p$}\xspace}
\newcommand{\pPb}{\mbox{$p$$+$Pb}\xspace}
\newcommand{\GeV}{\mbox{GeV/$c$}\xspace}

\newcommand{\PYTHIA}{{\sc pythia}\xspace}
\newcommand{\HIJING}{{\sc hijing}\xspace}
\newcommand{\GEANT}{{\sc geant}3\xspace}

\begin{document}

\title{Centrality-dependent modification of jet-production rates in
  deuteron-gold collisions at $\sqrt{s_{NN}}$=200~GeV}

\newcommand{\abilene}{Abilene Christian University, Abilene, Texas 79699, USA}
\newcommand{\augie}{Department of Physics, Augustana University, Sioux Falls, South Dakota 57197, USA}
\newcommand{\banaras}{Department of Physics, Banaras Hindu University, Varanasi 221005, India}
\newcommand{\barc}{Bhabha Atomic Research Centre, Bombay 400 085, India}
\newcommand{\baruch}{Baruch College, City University of New York, New York, New York, 10010 USA}
\newcommand{\bnlcoll}{Collider-Accelerator Department, Brookhaven National Laboratory, Upton, New York 11973-5000, USA}
\newcommand{\bnlphys}{Physics Department, Brookhaven National Laboratory, Upton, New York 11973-5000, USA}
\newcommand{\caucr}{University of California-Riverside, Riverside, California 92521, USA}
\newcommand{\charlesczech}{Charles University, Ovocn\'{y} trh 5, Praha 1, 116 36, Prague, Czech Republic}
\newcommand{\chonbuk}{Chonbuk National University, Jeonju, 561-756, Korea}
\newcommand{\ciae}{Science and Technology on Nuclear Data Laboratory, China Institute of Atomic Energy, Beijing 102413, People's Republic of~China}
\newcommand{\cns}{Center for Nuclear Study, Graduate School of Science, University of Tokyo, 7-3-1 Hongo, Bunkyo, Tokyo 113-0033, Japan}
\newcommand{\colorado}{University of Colorado, Boulder, Colorado 80309, USA}
\newcommand{\columbia}{Columbia University, New York, New York 10027 and Nevis Laboratories, Irvington, New York 10533, USA}
\newcommand{\czechtech}{Czech Technical University, Zikova 4, 166 36 Prague 6, Czech Republic}
\newcommand{\dapnia}{Dapnia, CEA Saclay, F-91191, Gif-sur-Yvette, France}
\newcommand{\elte}{ELTE, E{\"o}tv{\"o}s Lor{\'a}nd University, H-1117 Budapest, P{\'a}zm{\'a}ny P.~s.~1/A, Hungary}
\newcommand{\ewha}{Ewha Womans University, Seoul 120-750, Korea}
\newcommand{\fit}{Florida Institute of Technology, Melbourne, Florida 32901, USA}
\newcommand{\fsu}{Florida State University, Tallahassee, Florida 32306, USA}
\newcommand{\gsu}{Georgia State University, Atlanta, Georgia 30303, USA}
\newcommand{\hiroshima}{Hiroshima University, Kagamiyama, Higashi-Hiroshima 739-8526, Japan}
\newcommand{\howard}{Department of Physics and Astronomy, Howard University, Washington, DC 20059, USA}
\newcommand{\ihepprot}{IHEP Protvino, State Research Center of Russian Federation, Institute for High Energy Physics, Protvino, 142281, Russia}
\newcommand{\illuiuc}{University of Illinois at Urbana-Champaign, Urbana, Illinois 61801, USA}
\newcommand{\inrras}{Institute for Nuclear Research of the Russian Academy of Sciences, prospekt 60-letiya Oktyabrya 7a, Moscow 117312, Russia}
\newcommand{\instpasczech}{Institute of Physics, Academy of Sciences of the Czech Republic, Na Slovance 2, 182 21 Prague 8, Czech Republic}
\newcommand{\isu}{Iowa State University, Ames, Iowa 50011, USA}
\newcommand{\jaea}{Advanced Science Research Center, Japan Atomic Energy Agency, 2-4 Shirakata Shirane, Tokai-mura, Naka-gun, Ibaraki-ken 319-1195, Japan}
\newcommand{\jyvaskyla}{Helsinki Institute of Physics and University of Jyv{\"a}skyl{\"a}, P.O.Box 35, FI-40014 Jyv{\"a}skyl{\"a}, Finland}
\newcommand{\karoly}{K\'aroly R\'oberts University College, H-3200 Gy\"ngy\"os, M\'atrai \'ut 36, Hungary}
\newcommand{\kek}{KEK, High Energy Accelerator Research Organization, Tsukuba, Ibaraki 305-0801, Japan}
\newcommand{\korea}{Korea University, Seoul, 136-701, Korea}
\newcommand{\kurchatov}{National Research Center ``Kurchatov Institute", Moscow, 123098 Russia}
\newcommand{\kyoto}{Kyoto University, Kyoto 606-8502, Japan}
\newcommand{\labllr}{Laboratoire Leprince-Ringuet, Ecole Polytechnique, CNRS-IN2P3, Route de Saclay, F-91128, Palaiseau, France}
\newcommand{\lahorelums}{Physics Department, Lahore University of Management Sciences, Lahore 54792, Pakistan}
\newcommand{\lawllnl}{Lawrence Livermore National Laboratory, Livermore, California 94550, USA}
\newcommand{\losalamos}{Los Alamos National Laboratory, Los Alamos, New Mexico 87545, USA}
\newcommand{\lpc}{LPC, Universit{\'e} Blaise Pascal, CNRS-IN2P3, Clermont-Fd, 63177 Aubiere Cedex, France}
\newcommand{\lund}{Department of Physics, Lund University, Box 118, SE-221 00 Lund, Sweden}
\newcommand{\maryland}{University of Maryland, College Park, Maryland 20742, USA}
\newcommand{\mass}{Department of Physics, University of Massachusetts, Amherst, Massachusetts 01003-9337, USA}
\newcommand{\michigan}{Department of Physics, University of Michigan, Ann Arbor, Michigan 48109-1040, USA}
\newcommand{\muenster}{Institut f\"ur Kernphysik, University of Muenster, D-48149 Muenster, Germany}
\newcommand{\muhlenberg}{Muhlenberg College, Allentown, Pennsylvania 18104-5586, USA}
\newcommand{\myongji}{Myongji University, Yongin, Kyonggido 449-728, Korea}
\newcommand{\nagasaki}{Nagasaki Institute of Applied Science, Nagasaki-shi, Nagasaki 851-0193, Japan}
\newcommand{\nara}{Nara Women's University, Kita-uoya Nishi-machi Nara 630-8506, Japan}
\newcommand{\natmephi}{National Research Nuclear University, MEPhI, Moscow Engineering Physics Institute, Moscow, 115409, Russia}
\newcommand{\newmex}{University of New Mexico, Albuquerque, New Mexico 87131, USA}
\newcommand{\nmsu}{New Mexico State University, Las Cruces, New Mexico 88003, USA}
\newcommand{\ohio}{Department of Physics and Astronomy, Ohio University, Athens, Ohio 45701, USA}
\newcommand{\ornl}{Oak Ridge National Laboratory, Oak Ridge, Tennessee 37831, USA}
\newcommand{\orsay}{IPN-Orsay, Univ.~Paris-Sud, CNRS/IN2P3, Universit\'e Paris-Saclay, BP1, F-91406, Orsay, France}
\newcommand{\peking}{Peking University, Beijing 100871, People's Republic of~China}
\newcommand{\pnpi}{PNPI, Petersburg Nuclear Physics Institute, Gatchina, Leningrad region, 188300, Russia}
\newcommand{\riken}{RIKEN Nishina Center for Accelerator-Based Science, Wako, Saitama 351-0198, Japan}
\newcommand{\rikjrbrc}{RIKEN BNL Research Center, Brookhaven National Laboratory, Upton, New York 11973-5000, USA}
\newcommand{\rikkyo}{Physics Department, Rikkyo University, 3-34-1 Nishi-Ikebukuro, Toshima, Tokyo 171-8501, Japan}
\newcommand{\saispbstu}{Saint Petersburg State Polytechnic University, St.~Petersburg, 195251 Russia}
\newcommand{\saopaulo}{Universidade de S{\~a}o Paulo, Instituto de F\'{\i}sica, Caixa Postal 66318, S{\~a}o Paulo CEP05315-970, Brazil}
\newcommand{\seoulnat}{Department of Physics and Astronomy, Seoul National University, Seoul 151-742, Korea}
\newcommand{\stonybrkc}{Chemistry Department, Stony Brook University, SUNY, Stony Brook, New York 11794-3400, USA}
\newcommand{\stonycrkp}{Department of Physics and Astronomy, Stony Brook University, SUNY, Stony Brook, New York 11794-3800, USA}
\newcommand{\tenn}{University of Tennessee, Knoxville, Tennessee 37996, USA}
\newcommand{\titech}{Department of Physics, Tokyo Institute of Technology, Oh-okayama, Meguro, Tokyo 152-8551, Japan}
\newcommand{\tsukuba}{Center for Integrated Research in Fundamental Science and Engineering, University of Tsukuba, Tsukuba, Ibaraki 305, Japan}
\newcommand{\vandy}{Vanderbilt University, Nashville, Tennessee 37235, USA}
\newcommand{\waseda}{Waseda University, Advanced Research Institute for Science and Engineering, 17  Kikui-cho, Shinjuku-ku, Tokyo 162-0044, Japan}
\newcommand{\weizmann}{Weizmann Institute, Rehovot 76100, Israel}
\newcommand{\wigner}{Institute for Particle and Nuclear Physics, Wigner Research Centre for Physics, Hungarian Academy of Sciences (Wigner RCP, RMKI) H-1525 Budapest 114, POBox 49, Budapest, Hungary}
\newcommand{\yonsei}{Yonsei University, IPAP, Seoul 120-749, Korea}
\newcommand{\zagreb}{University of Zagreb, Faculty of Science, Department of Physics, Bijeni\v{c}ka 32, HR-10002 Zagreb, Croatia}
\affiliation{\abilene}
\affiliation{\augie}
\affiliation{\banaras}
\affiliation{\barc}
\affiliation{\baruch}
\affiliation{\bnlcoll}
\affiliation{\bnlphys}
\affiliation{\caucr}
\affiliation{\charlesczech}
\affiliation{\chonbuk}
\affiliation{\ciae}
\affiliation{\cns}
\affiliation{\colorado}
\affiliation{\columbia}
\affiliation{\czechtech}
\affiliation{\dapnia}
\affiliation{\elte}
\affiliation{\ewha}
\affiliation{\fit}
\affiliation{\fsu}
\affiliation{\gsu}
\affiliation{\hiroshima}
\affiliation{\howard}
\affiliation{\ihepprot}
\affiliation{\illuiuc}
\affiliation{\inrras}
\affiliation{\instpasczech}
\affiliation{\isu}
\affiliation{\jaea}
\affiliation{\jyvaskyla}
\affiliation{\karoly}
\affiliation{\kek}
\affiliation{\korea}
\affiliation{\kurchatov}
\affiliation{\kyoto}
\affiliation{\labllr}
\affiliation{\lahorelums}
\affiliation{\lawllnl}
\affiliation{\losalamos}
\affiliation{\lpc}
\affiliation{\lund}
\affiliation{\maryland}
\affiliation{\mass}
\affiliation{\michigan}
\affiliation{\muenster}
\affiliation{\muhlenberg}
\affiliation{\myongji}
\affiliation{\nagasaki}
\affiliation{\nara}
\affiliation{\natmephi}
\affiliation{\newmex}
\affiliation{\nmsu}
\affiliation{\ohio}
\affiliation{\ornl}
\affiliation{\orsay}
\affiliation{\peking}
\affiliation{\pnpi}
\affiliation{\riken}
\affiliation{\rikjrbrc}
\affiliation{\rikkyo}
\affiliation{\saispbstu}
\affiliation{\saopaulo}
\affiliation{\seoulnat}
\affiliation{\stonybrkc}
\affiliation{\stonycrkp}
\affiliation{\tenn}
\affiliation{\titech}
\affiliation{\tsukuba}
\affiliation{\vandy}
\affiliation{\waseda}
\affiliation{\weizmann}
\affiliation{\wigner}
\affiliation{\yonsei}
\affiliation{\zagreb}
\author{A.~Adare} \affiliation{\colorado} 
\author{C.~Aidala} \affiliation{\mass} \affiliation{\michigan} 
\author{N.N.~Ajitanand} \affiliation{\stonybrkc} 
\author{Y.~Akiba} \affiliation{\riken} \affiliation{\rikjrbrc} 
\author{H.~Al-Bataineh} \affiliation{\nmsu} 
\author{J.~Alexander} \affiliation{\stonybrkc} 
\author{M.~Alfred} \affiliation{\howard} 
\author{A.~Angerami} \affiliation{\columbia} 
\author{K.~Aoki} \affiliation{\kek} \affiliation{\kyoto} \affiliation{\riken} 
\author{N.~Apadula} \affiliation{\isu} \affiliation{\stonycrkp} 
% ...
\collaboration{PHENIX Collaboration} \noaffiliation

\date{\today}

%------------------------------------------------------------------------------|

\begin{abstract}

%\linenumbers

Jet production rates are measured in $p$$+$$p$ and $d$$+$Au 
collisions at $\sqrt{s_{NN}}$=200~GeV recorded in 2008 with the PHENIX 
detector at the Relativistic Heavy Ion Collider. Jets are reconstructed 
using the $R=0.3$ anti-$k_{t}$ algorithm from energy deposits in the 
electromagnetic calorimeter and charged tracks in multi-wire proportional 
chambers, and the jet transverse momentum ($p_T$) spectra are corrected 
for the detector response.  Spectra are reported for jets with 
$12<p_T<50$~GeV/$c$, within a pseudorapidity acceptance of 
$\left|\eta\right|<0.3$. The nuclear-modification factor ($R_{d{\rm Au}}$) 
values for 0\%--100\% $d$$+$Au events are found to be consistent with 
unity, constraining the role of initial state effects on jet production. 
However, the centrality-selected $R_{d{\rm Au}}$ values and 
central-to-peripheral ratios ($R_{\rm CP}$) show large, $p_T$-dependent 
deviations from unity, challenging the conventional models that relate 
hard-process rates and soft-particle production in collisions involving 
nuclei.

\end{abstract}

	
\maketitle

%\textbf{*** page break for PRL word count ***}  \clearpage

%%%%%%%%%%%%%%%%%%%%%%%%%%%%%%%%%%%%%%%%%%%%%%%%%%  intro

Jet cross-section measurements in \dAu collisions at the Relativistic 
Heavy Ion Collider (RHIC) are crucial for benchmarking the effects of the 
so-called cold-nuclear-matter environment, where jet production rates are 
expected to be sensitive to the modification of the nuclear parton 
densities~\cite{Eskola:2009uj} or to the energy loss of fast partons in 
the nucleus~\cite{Vitev:2007ve,Kang:2012kc,Kang:2015mta}. Recent 
observations of collective behavior in small collision systems at the 
Large Hadron Collider (LHC) and 
RHIC~\cite{CMS:2012qk,Aad:2012gla,Abelev:2012ola,Adare:2013piz} 
suggest that jet quenching in a possibly formed quark-gluon 
plasma~\cite{Zhang:2013oca} may play a role as well. Measurements of jet 
production as a function of centrality, an experimental proxy for the 
impact parameter of the deuteron with respect to the nucleus, are 
particularly important. They may reveal the impact parameter dependence of 
the nuclear parton densities~\cite{Helenius:2012wd}, of nonlinear quantum 
chromodynamics (QCD) effects at very high parton 
densities~\cite{Albacete:2012xq,Rezaeian:2012ye}, or of energy loss. More 
generally, they test the applicability of geometric models that describe 
how soft observables and hard process rates in heavy ion collisions are 
related~\cite{Miller:2007ri}. At RHIC energies, jet spectra have 
previously been reported only in \pp 
collisions~\cite{Abelev:2006uq,Bland:2013pkt}.

Modifications to jet production rates from the vacuum expectation are
quantified through the nuclear-modification factor $\RdAu \equiv
\left.  \left({\rm d}N^{\rm cent}/{\rm d}\pT\right) \right/ \left(
T_{d{\rm Au}}^{\rm cent} {\rm d}\sigma/{\rm d}\pT\right)$, where the
numerator is the per-event jet yield as a function of transverse
momentum (\pT) in a given class of \dAu collisions (``cent''), and the
denominator is the jet production cross section in \pp collisions
scaled by the corresponding mean value of the nuclear-overlap function
$T_{d{\rm Au}}$. Because $T_{d{\rm Au}}$ cannot be directly determined
experimentally, it is typically calculated within a Glauber model of
relativistic nuclear collisions. $\RdAu$ values of unity mean that the
jet rate in \dAu collisions is consistent with that in \pp collisions
after correcting for the larger degree of partonic overlap. The double
ratio of the \RdAu in central (large $T_{d{\rm Au}}$) events to that
in peripheral (small $T_{d{\rm Au}}$) events, \RCP, quantifies the
relative modification between \dAu event classes.

Previous measurements of hadron production at midrapidity in \dAu 
collisions~\cite{Adler:2006wg,Abelev:2009hx} found that \RdAu is 
consistent with unity at \pT=5--10~\GeV for all centralities, implying that 
hard-process yields scale with the overlap of the incoming partons and 
constraining the role of nuclear effects. The data further suggested that 
\RdAu for \pT$>$10~\GeV deviates from unity~\cite{Adler:2006wg}, but with 
small statistical significance. Recent measurements of 
\pT$\gtrsim$100~\GeV jet and dijet production in \pPb collisions at the 
LHC showed a large, unexpected sensitivity to the collision 
centrality~\cite{ATLAS:2014cpa,Chatrchyan:2014hqa}. A number of novel 
explanations~\cite{Bzdak:2014rca,Alvioli:2014eda,Armesto:2015kwa} have 
been proposed for these effects, which are generally expected to persist 
to RHIC energies, but at large \pT where previous measurements have lacked 
statistical precision. This Letter presents the centrality dependence of 
jet production in an asymmetric collision system over a kinematic range 
previously not measured at RHIC.

%%%%%%%%%%%%%%%%%%%%%%%%%%%%%%%%%%%%%%%%%%%%%%%%%%%%%%%% PHENIX

Jets were measured in one of the PHENIX central spectrometers (the 
``East'' arm)~\cite{Adcox:2003zm} during data taking in 2008.  The 
spectrometer provides a pseudorapidity aperture of $\left|\eta\right| 
< 0.35$, $\pi/2$ coverage in azimuth, and is situated outside a 
$0.9$~T axial magnetic field. Charged-particle tracks are measured by 
a set of multi-wire proportional chambers, including an inner drift 
chamber and multiple outer pad chambers that together provide a 
resolution of $\sigma_{p}/p = 0.7\% \oplus 1\% p$ where $p$ is in 
\GeV. Energy deposits from neutral particles are measured by the 
finely segmented electromagnetic calorimeter, composed of two 
lead-glass \v{C}erenkov and two lead-scintillator sectors, which have 
a resolution determined by beam tests~\cite{David:1998sk} to be 
$\sigma_{E}/E = 5.9\%/\sqrt{E} \oplus 0.8\%$ and $8.1\%/\sqrt{E} 
\oplus 2.1\%$, respectively, where $E$ is in~GeV.  Calibration was 
performed through the reconstruction of neutral pion decays. The 
calorimeter further provides a trigger signal initiated by the 
presence of at least $1.6$ or $2.1$~GeV of energy deposited in one of 
the groups of overlapping $4\times4$ towers in the lead-glass or 
lead-scintillator modules, respectively. In addition to the 
spectrometer, a pair of beam--beam counter detectors situated along 
the beam line at $3.0 < \left|\eta\right| < 3.9$ provide the 
minimum-bias trigger signal and reconstruct the $z$ position of the 
primary vertex.

The analyzed \pp and \dAu data sets were carefully chosen, and the
single central arm was used, to ensure a large, stable and uniform
acceptance for jets, and corresponded to $2.0$~pb$^{-1}$ and
$23$~nb$^{-1}$ (equivalent to an integrated nucleon--nucleon
luminosity of $9.1$~pb$^{-1}$), respectively. The centrality of \dAu
collisions was characterized using the total charge deposited in the
Au-going beam-beam counter. A Glauber Monte
Carlo~\cite{Miller:2007ri,Alver:2008aq} description of \dAu collisions
was used, along with the hypothesis that this charge increased
linearly with the number of nucleon--nucleon
collisions~\cite{Adare:2013nff}, to determine the fraction of \dAu
collisions accepted by the minimum-bias trigger, $88\pm4$\%, and to
estimate the mean value of the nuclear-overlap function 
$T_{d{\rm Au}}^{\rm cent}$ for 0\%--100\% centrality events, as well 
as those defined by the centrality intervals (``cent'') of 0\%--20\%,
20\%--40\%, 40\%--60\%, and 60\%--88\%. The relationship between the
Au-going charge and the collision geometry has been validated through,
for example, an analysis of forward neutron production in \dAu
collisions, and analyses of \pp collisions indicate that it should
hold for events that produce $\pT=20$~GeV hadrons~\cite{Adare:2013nff}.

%%%%%%%%%%%%%%%%%%%%%%%%%%%%%%%%%%%%%%%%%%%%%%%%% basic measurement

In this analysis, the final-state jet definition is specified by
applying the anti-$k_{t}$
algorithm~\cite{Cacciari:2008gp,Cacciari:2011ma} with radius parameter
$R=0.3$ to electromagnetic clusters (in the calorimeter) and
charged-particle tracks (in the drift and pad chambers), each with a
minimum $\pT$ of $0.4$~\GeV.  The anti-$k_{t}$ algorithm clusters
outward from the hard core of jets, reducing the sensitivity to
detector edges. A detailed set of criteria designed to select charged
particles with a well-measured momentum while ensuring a large and
uniform acceptance were applied to candidate reconstructed
tracks. Clusters consistent with arising from the same particle as a
reconstructed track were rejected to avoid double counting jet
constituent energy.  Jets which are dominated by reconstructed tracks
with a large, erroneously measured \pT~\cite{Adcox:2002pe} were
rejected by requiring at least three constituent particles and by
requiring at least one quarter of the momentum to arise from
clusters. To ensure that the core of the jet is fully contained within
the detector, the jet axis was required to be separated from the edge
of the acceptance by $0.05$ units in pseudorapidity and azimuth.

Detector-level jets, defined as those passing the above criteria, were
used to form a transverse momentum spectrum (\pTrec) in each event
class. The contribution of the small underlying event background was
not subtracted on a jet-by-jet basis, but was corrected for in the
unfolding procedure described below. Jets were selected from the
triggered data if a jet constituent fell into the same region of the
calorimeter that provided the trigger signal. The trigger efficiency
was estimated for each event class by checking this condition as a
function of \pTrec in minimum-bias events. The \pTrec-level spectra
were corrected for this efficiency, which rose monotonically with
\pTrec and was approximately $70$\% ($98$\%) at $10$~\GeV ($25$~\GeV).

%%%%%%%%%%%%%%%%%%%%%%%%%%%%%%%%%%%%%%%%%%%%%%%%%% MC 

Monte Carlo simulations were used to determine the response of the 
detector to jets and to correct the measured spectra. In simulation, jets 
are defined by applying the anti-$k_{t}$ algorithm to long-lived primary 
particles, resulting in jets with a particle-level transverse momentum 
(\pT). The \PYTHIA 6.4 event generator~\cite{Sjostrand:2006za} with the 
\textsc{d}{\footnotesize 6}\textsc{t}~tune~\cite{Albrow:2006rt} and 
\textsc{cteql}{\footnotesize 1} parton 
distribution 
function set~\cite{Pumplin:2002vw} was used to generate hard scattering 
\pp events with a jet within the acceptance of the East arm. Six separate 
samples with exclusive selections on the hard-scattering momentum transfer 
in \PYTHIA, consisting of $10^{5}$ events each, were weighted according to their partial cross-section and combined to form a \pT spectrum from $8$ to $80$~\GeV. The response of 
the detector was simulated with \GEANT~\cite{Brun:1987ma} and the 
resulting events were analyzed identically to the data. To understand the 
effects of the underlying event in \dAu collisions, jet reconstruction was 
also performed on the simulated events after they were embedded into 
minimum-bias \dAu data events of each centrality. In each event class, 
particle-level jets were matched with detector-level jets and the 
correspondence between the true \pT and the measured \pTrec was collected 
into a response matrix $\mathcal{R}(\pT, \pTrec)$.

The reconstruction and selection efficiency, $\epsilon(\pT)$, for 
particle-level jets within $\left|\eta\right| < 0.3$ rose with \pT and was 
$\approx35$\% ($50$\%) at $10$~\GeV ($25$~\GeV) in \pp collisions. The 
inefficiency was dominated by the minimum requirement on the calorimetric 
fraction of the jet momentum. For a given selection on the particle-level 
jet \pT, the mean value of the $\pTrec/\pT$ distribution 
$\approx0.65$-$0.70$ resulted from missing neutral hadronic energy and 
tracking inefficiency. The width of this distribution was 
$\approx$20\%--25\%, rose slightly with \pT, and was driven by 
jet-by-jet 
fluctuations in the neutral hadronic momentum fraction and not by the 
resolution on the constituent momenta. In the \dAu event classes, the 
impact of the underlying event on the response decreased systematically 
with increasing jet \pT. For \pT=20~\GeV jets in 0\%--20\% centrality 
\dAu\ events, the underlying event background increased the efficiency by 
2\%, the average \pTrec by 0.1--0.2~\GeV, and the \pTrec resolution by 
1\%, relative to that in \pp\ events.

%%%%%%%%%%%%%%%%%%%%%%%%%%%%%%%%%%%%%%%%%%%%%%%%%%% corrections

The \pTrec-level spectra were corrected for the detector response and the 
presence of the underlying event in \dAu collisions through the 
singular-value-decomposition unfolding 
method~\cite{Hocker:1995kb,Adye:2011gm}.  For an observed spectrum ${\rm 
d}N/{\rm d}\pTrec$, this method inverts the equation ${\rm d}N/{\rm 
d}\pTrec =\mathcal{R}\cdot{\rm d}N/{\rm d}\pT$ by expressing ${\rm 
d}N/{\rm d}\pT$ as a linear combination of the left singular vectors of 
$\mathcal{R}$, with coefficients determined by ${\rm d}N/{\rm d}\pTrec$. 
This inversion is regularized by keeping the contribution only from the 
$k$ vectors with the largest singular values.  The contribution from the 
remaining vectors is truncated to ensure that ${\rm d}N/{\rm d}\pT$ is 
unaffected by statistical fluctuations.

Following standard techniques~\cite{Hocker:1995kb}, $k$ was fixed at 
$5$, and the results were validated by comparing ${\rm d}N/{\rm d}\pT$, 
propagated through $\mathcal{R}$, to ${\rm d}N/{\rm d}\pTrec$, and by 
examining the curvature of ${\rm d}N/{\rm d}\pT$ with respect to the 
simulated \pT spectrum used to populate $\mathcal{R}$. The iterative 
Bayesian method~\cite{D'Agostini:1994zf} gave consistent results.  The 
statistical uncertainties on ${\rm d}N/{\rm d}\pT$ were evaluated by 
resampling ${\rm d}N/{\rm d}\pTrec$ according to its uncertainties and 
observing the changes in ${\rm d}N/{\rm d}\pT$. Finally, the 
${\rm d}N/{\rm d}\pT$ spectra were corrected for the reconstruction 
efficiency $\epsilon(\pT)$. At low~\pT in 0\%--20\% events, the \RdAu 
after unfolding was lower than the detector-level \RdAu by $\approx$20\%, 
while the two are comparable at high~\pT or in peripheral events.

\begin{figure}[!htb]
\includegraphics[width=1.0\linewidth]{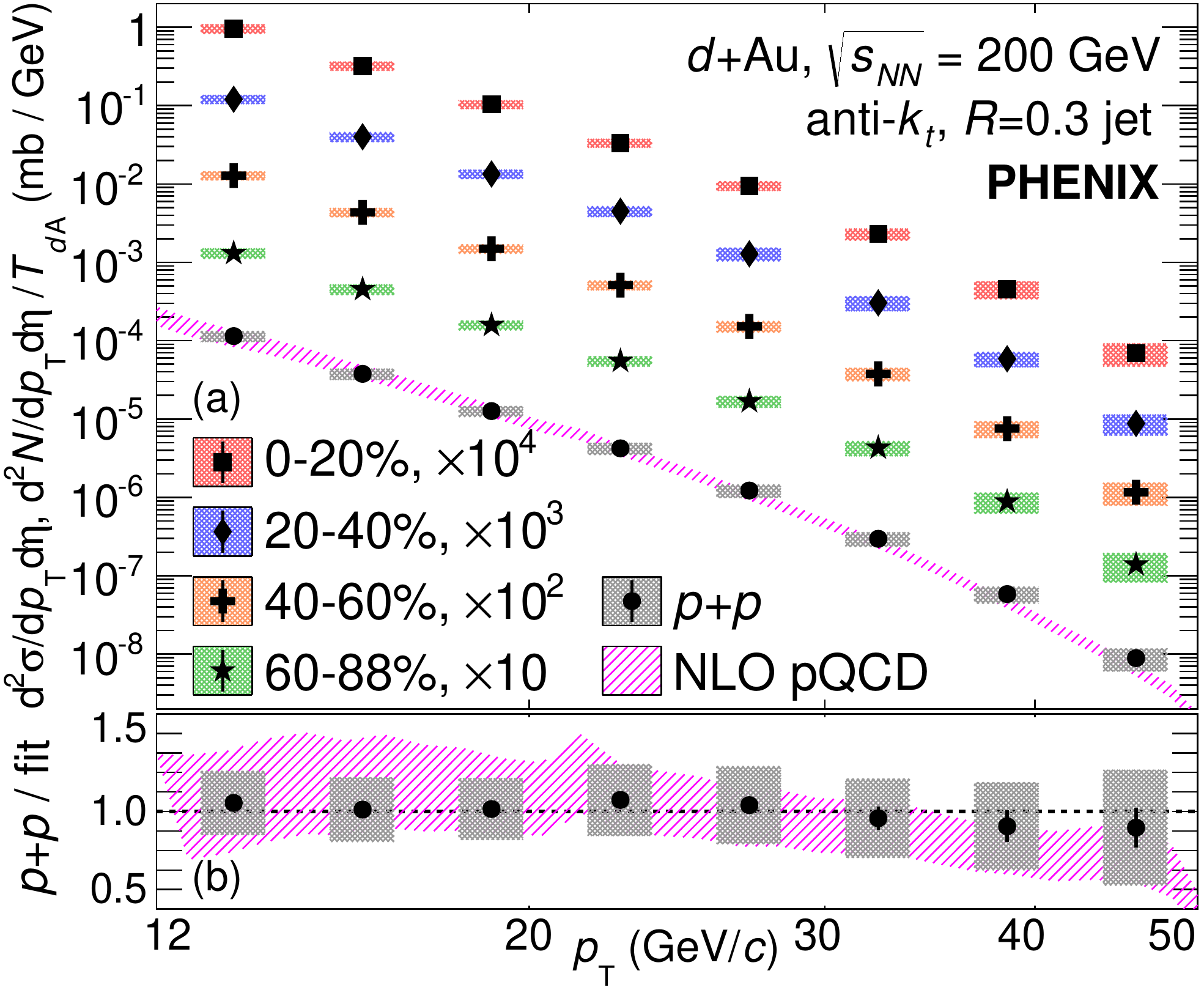}
\caption{\label{fig:fig1} (Color online)
Measured anti-$k_{t}$, $R=0.3$ jet yields in \dAu collisions, and the 
measured and calculated jet cross section in \pp collisions, with the data 
series offset by multiplicative factors. Total systematic uncertainties, 
including overall normalization uncertainties, and statistical 
uncertainties are shown as shaded bands and vertical bars, respectively. 
In the bottom panel, the \pp data and perturbative QCD 
calculation~\protect\cite{Nagy:2003tz,Ball:2010de} are divided by a fit 
to the data.
}
\end{figure}

The \pp differential cross section was constructed~\cite{Adler:2006wg} via 
$2\pi\sigma^{pp}N^{\rm jet}(\pT)/\epsilon^{pp}N^{\rm evt}\epsilon(\pT)\Delta\pT\Delta\eta\Delta\phi$, 
where $\sigma^{pp}=23.0\pm2.2$~mb is the minimum-bias cross section, 
$\epsilon^{pp}=0.79\pm0.02$ is the fraction of jet events meeting the 
minimum-bias condition, and $2\pi/\Delta\pT\Delta\eta\Delta\phi$ are 
phase-space factors.  Figure~\ref{fig:fig1} shows the \dAu yields and the 
\pp cross section, which compares well with a 
perturbative QCD calculation~\cite{Nagy:2003tz,Ball:2010de}.

%-------------------------------------------------- Fig_1

%-------------------------------------------------- Fig_2
\begin{figure}[!htb]
\includegraphics[width=1.0\linewidth]{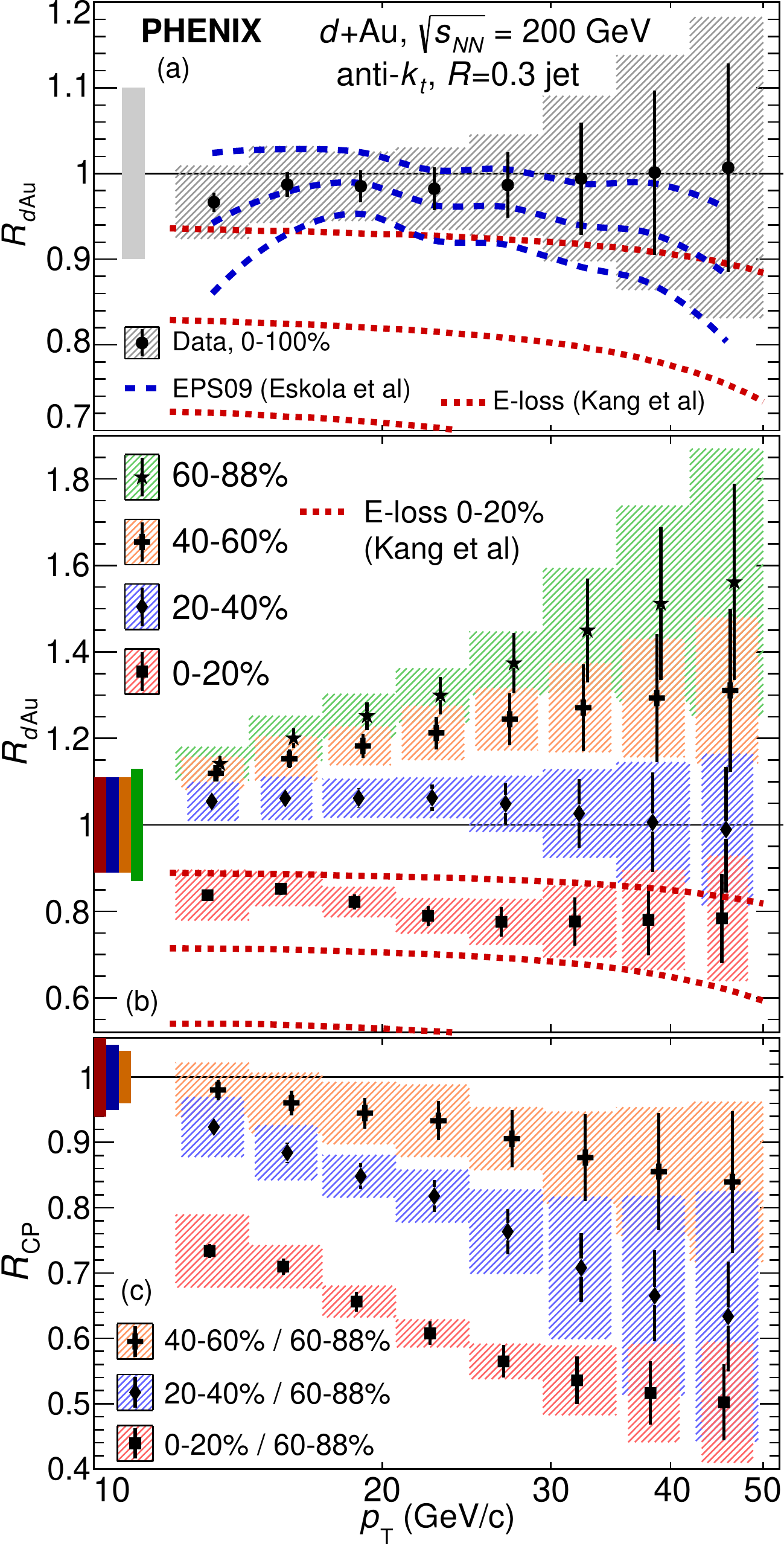}
\caption{\label{fig:fig2} (Color online)
\RdAu for (a) 0\%--100\% and (b) centrality-selected collisions, and (c) 
\RCP, as a function of \pT. Systematic, statistical and normalization 
uncertainties are shown as shaded bands, vertical bars, and the leftmost 
bands centered at $1$, respectively. When error bands overlap vertically, 
their horizontal widths have been adjusted so that both are visible. 
Dashed lines show the uncertainty range of calculations incorporating 
nuclear parton densities~\cite{Eskola:2009uj} and energy 
loss~\cite{Kang:2015mta}.
}
\end{figure}

%%%%%%%%%%%%%%%%%%%%%%%%%%%%%%%%%%%%%%%%%%%%%%%%%%%%% systematics

The measured spectra and nuclear-modification factors are subject to 
systematic uncertainties from a variety of sources. For most sources, the 
effects on the results were determined by modifying the simulation sample, 
the event or jet-selection criteria, or the unfolding procedure itself, 
and repeating the analysis. The variations were applied simultaneously in 
the analyses of the \dAu and \pp spectra to allow for their full or 
partial cancellation in the \RdAu and \RCP quantities, with the exception 
of the variation of $k$, described below.

The impact of uncertainties on the detector energy scales was determined 
by varying the momenta of the reconstructed tracks and clusters in 
simulation. The cluster energies were varied by $3$\%. The track momenta 
were varied by a track \pT-dependent amount, which was $2$\% for $\pT \le 
10$~\GeV and increased linearly to $4$\% for $\pT = 30$~\GeV. The 
sensitivity of the results to the jet selection was evaluated by varying 
the maximum and minimum requirement on the calorimetric content of the 
jet, and by raising the required number of jet constituents. The 
uncertainty in the jet acceptance was evaluated by doubling the fiducial 
distance between jets and the edges of the detector, and by restricting 
the vertex $z$ position to a narrower range. The uncertainties associated 
with the unfolding procedure were evaluated by changing the power law 
index of the simulated \pT spectrum by $\pm1$, and by increasing and 
decreasing the value of $k$. Because they are statistical in nature, the 
effects on the spectra from varying $k$ were treated as uncorrelated 
between the event classes. The sensitivity to the underlying physics model 
was evaluated by performing the corrections with a sample of \PYTHIA 
events analogous to the nominal one but generated with 
{\sc tune~a}~\cite{Field:2002vt} and the 
{\sc cteq{\footnotesize 5}l}~\cite{Lai:1999wy} set. A 2\% 
uncertainty, uncorrelated between event classes, was assigned to the 
spectra below $25$~\GeV to cover possible defects in modeling the trigger 
efficiency.

For each observable, the magnitudes of the resulting changes were added in 
quadrature to obtain a total systematic uncertainty. The total uncertainty 
on the spectra increased from $12$\% at $\pT = 12$~\GeV to $30$\% or 
higher at $\pT = 50$~\GeV and was dominated at all \pT by the energy 
scale. Because the reconstruction procedure in \dAu and \pp collisions was 
identical, and the performance, corrections and resulting spectra are very 
similar, the effects of the variations on \RdAu and \RCP canceled to a 
large degree. The uncertainties on this quantity ranged from $4$\% at $\pT 
= 12$~\GeV (with no single source dominating) to $15$\% or higher 
(dominated by unfolding and physics model) at $\pT = 50$~\GeV.

Additional normalization uncertainties on the \pp cross section of $10$\% 
arose from the uncertainty on $\sigma^{pp}/\epsilon^{pp}$. Uncertainties 
in the determination of $T_{d{\rm Au}}$ contributed to the \RdAu and \RCP, 
such that the total uncertainty on these ranged from $3$\% to $13$\%.

% results

Figure~\ref{fig:fig2} summarizes the measured \RdAu and \RCP
quantities.  The 0\%--100\% \RdAu is consistent with unity at all \pT
values and is \pT-independent within uncertainties. The data are
consistent with a next-to-leading order
calculation~\cite{NestorArmesto,Frixione:1995ms,Frixione:1997np,Frixione:1997ks}
incorporating the EPS09~\cite{Eskola:2009uj} nuclear-parton-density
set, suggesting that nuclear effects are small at high-$Q^2$ in the
nuclear Bjorken-$x$ range $\approx 0.1$--$0.5$. When compared to calculations
over a range of energy loss rates in the cold
nucleus~\cite{Kang:2015mta}, the data favor only small momentum
transfers between the hard-scattered parton and nuclear material,
providing constraints on initial-state, or any additional final-state,
energy loss.

In contrast, the centrality-dependent \RdAu values strongly deviate
from unity, manifesting as a suppression ($\RdAu < 1$) and enhancement
($\RdAu > 1$) in central and peripheral collisions respectively, which
increase in magnitude with \pT.  Accordingly, the \RCP is $<1$ in most
selections and decreases systematically with \pT and in more central
events. While the suppressed \RdAu in 0\%--20\% events is consistent
with a calculation incorporating modest energy loss, an enhancement in
40\%--88\% events, which coincidentally cancels with the suppression to
produce an unmodified minimum bias rate, is challenging to understand
as a distinct physics effect.

If jet production is unmodified but a physics bias enters into the
centrality classification, this could naturally explain the \RdAu
results. In fact, measurements of centrality-dependent yields are
understood to be biased by the increased multiplicity in
hard-scattering nucleon-nucleon
events~\cite{Adler:2006xd,Adare:2013nff,Perepelitsa:2014yta,Adam:2014qja},
which generally increases (decreases) the yield in central
(peripheral) collisions. The results have been corrected for this bias
following Ref.~\cite{Adare:2013nff}, thus slightly increasing the
magnitude of the modifications.   On the other hand, if the charged
particle multiplicity several units of rapidity away in the Au-going
direction were suppressed instead of enhanced in $\pT > 12$~\GeV\ jet
events, this would reverse the sign of the correction and could result
in the observed modifications.  The jet \pT-dependence of this
correlation has been studied in \pp data and in
\HIJING~\cite{Gyulassy:1994ew}, where it is well-reproduced. The
decreased multiplicity results in modest changes ($< 5$\%) in the
correction factors for events with $\pT =
20$~\GeV\ hadrons~\cite{Adare:2013nff}, a much smaller effect than
what is needed to describe the \RdAu data. Thus, no feature of
elemental \pp collisions can explain the data alone, indicating the
relevance of the large nucleus and the need for successful models to
describe the correlation between soft and hard processes in \pp and
\dAu.

At midrapidity, jet production in \pPb collisions at the
LHC~\cite{ATLAS:2014cpa} follows a similar modification pattern in 
the Bjorken-$x$ range, $x_p \sim x_\mathrm{Pb} \gtrsim 0.1$. However,
the $R_{p\mathrm{Pb}}$ in those results scales with proton-$x$,
suggesting a scenario in which the modifications arise from a novel
feature of the proton wavefunction at large
$x$~\cite{Bzdak:2014rca,Alvioli:2014eda,Armesto:2015kwa}. For example,
if high-$x$ deuteron configurations have a weaker than average
interaction strength and strike fewer nucleons in the Au
nucleus~\cite{Alvioli:2014eda}, this would result in the unmodified,
suppressed and enhanced \RdAu in minimum-bias, central and peripheral
events, respectively. If so, the observed centrality dependence of
forward hadron
production~\cite{Arsene:2004ux,Adler:2004eh,Abelev:2007nt,Adare:2011sc}
in \dAu collisions may arise from the same mechanism as the results
presented here, because both are kinematically associated with the
scattering of a large-$x$ parton in the deuteron.  Finally, using an
alternate estimate of $T_{d{\rm Au}}$ provided by applying the
Glauber--Gribov color fluctuation
model~\cite{Alvioli:2013vk,Aad:2015zza} to the data would increase the
deviation of \RdAu in the most central and peripheral events from
unity by 10\% and 5\%, respectively.

This Letter presents the first measurement of high-\pT jet production
in \dAu collisions at RHIC. The jet rate in inclusive collisions is
broadly consistent with expectations, providing constraints in a new
kinematic regime on modifications to the parton densities in nuclei
and on the energy loss of fast partons in the nuclear medium. When
compared to the expectation from geometric considerations, the rates
in centrality-selected events strongly deviate from unity, featuring
suppression and enhancement patterns in central and peripheral events,
respectively.  These deviations grow with increasing \pT, but cancel
in the overall jet rate, and challenge the conventional pictures of
how hard-process rates and soft-particle production are related in
collisions involving nuclei.

%\textbf{*** page break for PRL word count $<$3.5 pages $<$7 columns}  
%\clearpage

%%%%%%%%%%%%%%%%%%%%%%%%%  Acknowledgments 

%\section*{ACKNOWLEDGMENTS}   % Run-14 long form for all journals

We thank the staff of the Collider-Accelerator and Physics
Departments at Brookhaven National Laboratory and the staff of
the other PHENIX participating institutions for their vital
contributions.  We acknowledge support from the
Office of Nuclear Physics in the
Office of Science of the Department of Energy,
the National Science Foundation,
Abilene Christian University Research Council,
Research Foundation of SUNY, and
Dean of the College of Arts and Sciences, Vanderbilt University
(U.S.A),
Ministry of Education, Culture, Sports, Science, and Technology
and the Japan Society for the Promotion of Science (Japan),
Conselho Nacional de Desenvolvimento Cient\'{\i}fico e
Tecnol{\'o}gico and Funda\c c{\~a}o de Amparo {\`a} Pesquisa do
Estado de S{\~a}o Paulo (Brazil),
Natural Science Foundation of China (People's Republic of~China),
Croatian Science Foundation and
Ministry of Science, Education, and Sports (Croatia),
Ministry of Education, Youth and Sports (Czech Republic),
Centre National de la Recherche Scientifique, Commissariat
{\`a} l'{\'E}nergie Atomique, and Institut National de Physique
Nucl{\'e}aire et de Physique des Particules (France),
Bundesministerium f\"ur Bildung und Forschung, Deutscher
Akademischer Austausch Dienst, and Alexander von Humboldt Stiftung 
(Germany),
National Science Fund, OTKA, K\'aroly R\'obert University College,
and the Ch. Simonyi Fund (Hungary),
Department of Atomic Energy and Department of Science and Technology 
(India),
Israel Science Foundation (Israel),
Basic Science Research Program through NRF of the Ministry of Education 
(Korea),
Physics Department, Lahore University of Management Sciences (Pakistan),
Ministry of Education and Science, Russian Academy of Sciences,
Federal Agency of Atomic Energy (Russia),
VR and Wallenberg Foundation (Sweden),
the U.S. Civilian Research and Development Foundation for the
Independent States of the Former Soviet Union,
the Hungarian American Enterprise Scholarship Fund,
and the US-Israel Binational Science Foundation.

%\bibliography{ppg184x2}   

 
\end{document}